\documentclass[twocolumn,prl,tightenlines,superscriptaddress,showpacs]{revtex4-2}

\usepackage{amsmath}
\usepackage{amssymb,amsfonts,latexsym}
\usepackage{bm}
\usepackage[mathcal]{euscript}
\usepackage{graphicx}
\usepackage{epsfig}
\usepackage[svgnames]{xcolor}
\usepackage{xfrac}
\usepackage{mathrsfs}

\usepackage[colorlinks,linkcolor=MediumBlue,citecolor=MediumBlue,urlcolor = MediumBlue]{hyperref}
\usepackage{url}

\newcommand{\colrev}[1]{{\color{black} #1}}

\begin{document}

\title{Dynamical pattern formation without self-attraction in quorum-sensing active matter: \\ the interplay between nonreciprocity and motility}

\author{Yu Duan}
\affiliation{Max Planck Institute for Dynamics and Self-Organization (MPI-DS), 37077 G\"ottingen, Germany}

\author{Jaime Agudo-Canalejo}
\affiliation{Max Planck Institute for Dynamics and Self-Organization (MPI-DS), 37077 G\"ottingen, Germany}

\author{Ramin Golestanian} \email{ramin.golestanian@ds.mpg.de}
\affiliation{Max Planck Institute for Dynamics and Self-Organization (MPI-DS), 37077 G\"ottingen, Germany}
\affiliation{Rudolf Peierls Centre for Theoretical Physics, University of Oxford, Oxford OX1 3PU,
United Kingdom}

\author{Beno\^{\i}t Mahault} \email{benoit.mahault@ds.mpg.de}
\affiliation{Max Planck Institute for Dynamics and Self-Organization (MPI-DS), 37077 G\"ottingen, Germany}

\date{\today}

\begin{abstract}

We study a minimal model involving two species of particles interacting via quorum-sensing rules. 
Combining simulations of the microscopic model and linear stability analysis of the associated coarse-grained field theory, 
we identify a mechanism for dynamical pattern formation that
does not rely on the standard route of intra-species effective attractive interactions. 
Instead, our results reveal a highly dynamical phase of chasing bands induced only by the combined effects of self-propulsion and nonreciprocity in the inter-species couplings.
Turning on self-attraction, we find that the system may phase separate into a macroscopic domain of such chaotic chasing bands coexisting with a dilute gas. 
We show that the chaotic dynamics of bands at the interfaces of this phase-separated phase results in anomalously slow coarsening.

\end{abstract}

\maketitle


 Active systems are driven out-of-equilibrium at the level of their microscopic constituents.
 Since activity may arise in various forms, including particle motility~\cite{Romanczuk2012ABPs}, local generation of nonconservative forces~\cite{Prost2015ActGels,Doostmohammadi2018NatCom} and torques~\cite{uchida2010,Furthauer2012chiral}, 
 growth~\cite{DellArciprete2018NatCom}, or sustained chemical reactions~\cite{Golestanian2019phoretic},
 the active matter field is rapidly expanding in multiple directions~\cite{Gompper2020JPCM}.
 Recently, the generation of interactions breaking action-reaction symmetry~\cite{ivlev2015statistical,saha2020scalar,you2020nonreciprocity,fruchart2021non,GelbwaserPRL2021} 
 as a paradigm for activity has received increasing attention.
 Such nonreciprocal interactions occur in a variety of contexts, including particles interacting through a nonequilibrium medium~\cite{Dzubiella2003PRL,SotoPRL2014,wysocki2016propagating,wittkowski2017nonequilibrium,agudo2019active,Nasouri2020PRL,banerjee2022unjamming,gupta2022active}, or via social forces~\cite{chen2017fore,bauerle2018self,gomez2022intermittent}.
 
 Due to connections built with non-Hermitian physics~\cite{Ganainy2018,Shankar2022NatPhys}, nonreciprocity has been argued to constitute a generic route for the emergence of steady states breaking time reversal symmetry (TRS)~\cite{agudo2019active,saha2020scalar,you2020nonreciprocity,frohoff2021suppression,fruchart2021non,Ouazan2021EPJE,Loos2022,Saha2022}.
 Examples include rotating chiral phases in flocking systems involving multiple species~\cite{fruchart2021non}, or traveling patterns in phase-separating mixtures~\cite{saha2020scalar,you2020nonreciprocity}. 
 The latter are described by the nonreciprocal Cahn-Hilliard (NRCH) model~\cite{saha2020scalar,you2020nonreciprocity}, which generalizes the Cahn-Hilliard theory of phase separation~\cite{CH} by including 
 a nonequilibrium chemical potential with
 antisymmetric couplings between species.
Importantly, the NRCH model predicts that preconditions for the emergence of TRS broken phases are the presence of intra-species attraction that drives demixing, as well as chasing inter-species interactions~\cite{saha2020scalar,you2020nonreciprocity}.
 In addition to dynamical patterns, the similarity between the NRCH model and reaction-diffusion equations~\cite{diego2018key,haas2021turing} has led to additional connections such as the presence of Turing-like instabilities~\cite{frohoff2021suppression,frohoff2022nonreciprocity}.
 
  
 \colrev{
 A model mechanism for self-organization in motile active matter are quorum-sensing interactions.
 Quorum-sensing mediated by chemical signals is for example known to drive swarming and pattern formation in bacterial populations~\cite{Daniels2004QS,Liu2011Science,curatolo2020cooperative}.
 Collective aggregation was moreover shown to arise in suspensions of light-activated colloids whose motility is locally regulated by 
 their density via feedback control loops~\cite{Gomez-Solano2017,bauerle2018self,lavergne2019Science}.
 Often, quorum-sensing interactions are modelled as a direct response of particle motility to local density variations~\cite{cates2015motility,Solon2018NJP,Fischer2020PRE}.
 Despite their simplicity, minimal quorum-sensing models are able to capture complex collective behaviors~\cite{bauerle2018self,curatolo2020cooperative}.
 A notable example is the motility induced phase separation (MIPS), whose origin is rooted in the effective attraction induced by the self-inhibition of motility~\cite{cates2015motility}.
 In addition, nonreciprocal couplings naturally arise when multiple species exhibit quorum-sensing with asymmetric responses~\cite{wittkowski2017nonequilibrium,dinelli2022self},
 thus without resorting to explicit nonreciprocal pairwise forces~\cite{chiu2023phase,mandal2022robustness,Kreienkamp_2022}.
 For strong nonreciprocity and in the phase separation regime, related models were shown to lead to TRS broken phases in qualitative agreement with the picture provided by the NRCH model~\cite{wysocki2016propagating,wittkowski2017nonequilibrium,dinelli2022self}.
}

  In this Letter, we study a binary model of quorum-sensing self-propelled particles in two dimensions.
  Our simulations reveal the presence of thin traveling bands even in the absence of intra-species couplings.
  Instead, and as confirmed by the associated coarse-grained field theory, they arise from a new mechanism relying only on self-propulsion and chasing interactions.
  In large systems, these bands self-organize into a spatio-temporal chaotic phase
  which we find to be involved in phase-separated configurations at moderate self-attraction.
  We show that this phase-separated phase coarsens with a scaling exponent distinct from that of standard MIPS~\cite{Stenhammar2014SoftMatt,caporusso2020motility,shi2020self,caporusso2022dynamics}.

\begin{figure*}[t!]
	\includegraphics[width=1.0\linewidth]{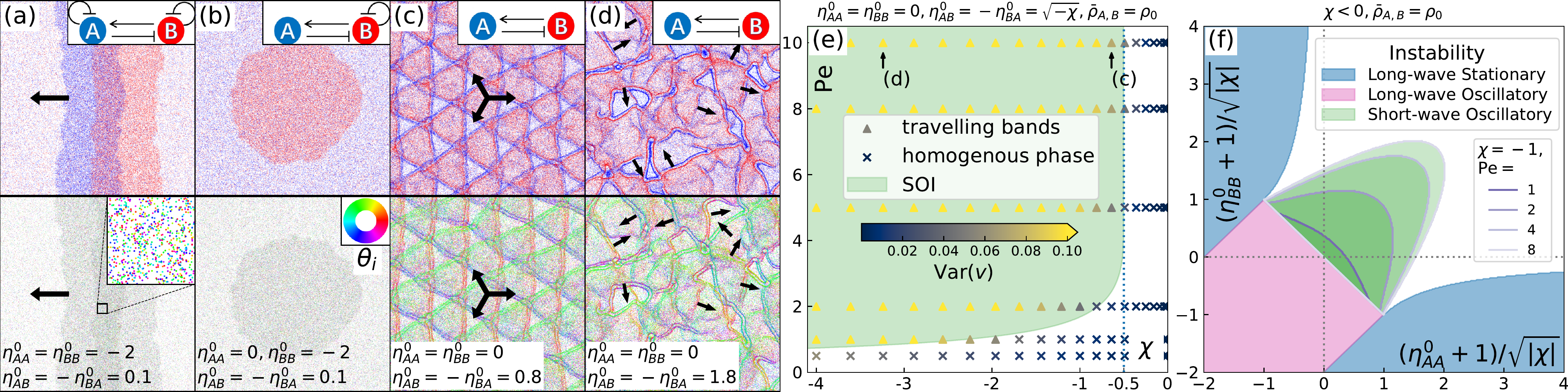}
	\caption{
	(a-d) Representative simulation snapshots for strong (a,b) and vanishing (c,d) motility self-inhibition in the presence of chasing interactions ($\chi < 0$).
	The top (bottom) row shows the particles color-coded by their species (polarity orientation), while the thick arrows indicate the travelling direction of the patterns.
	\colrev{The zoom in (a) illustrates how disordered configurations appear in grey.}
	The top captions give a graphical representation of the interaction rules with arrows (bars) denoting motility activation (inhibition).
	Parameters: $\bar{\rho}_{A,B}=\rho_0=80$, $\mathrm{Pe}=10$ and $L_x=L_y=40$.
	(e) ${\rm Var}(v)$ (see definition in the text) as function of $\chi$ and ${\rm Pe}$ in the vanishing self-inhibition regime, triangles and crosses indicate points where traveling bands and the homogeneous phase are observed. 
	The green shaded region marks the domain of existence of SOI.
	(f) Typical linear stability diagram
	 for $\chi<0$ and $\bar{\rho}_{A,B}=\rho_0$.
	 }
	\label{FIG1}
\end{figure*}

\paragraph{Description of the model.---}
We consider a dynamics where 
the position $\mathbf{r}_{i}$ and orientation $\hat{\mathbf{u}}_{i}=(\cos \theta_{i},\sin\theta_{i})$ of particle $i$ from species $S\in\{A, B\}$ evolve at time $t$ according to
\begin{equation} \label{eq:micro-model}
    \dot{\mathbf{r}}_{i,S} = v_S\left[\tilde{{\rho}}_A, \tilde{\rho}_B\right] \hat{\mathbf{u}}_{i,S} , \qquad
    \dot{\theta}_{i,S} = \sqrt{2D_r}\xi_i(t).
\end{equation}
$\xi_i$ is a Gaussian white noise with zero mean and unit variance, while $D_r$ denotes the corresponding rotational diffusivity
assumed equal for the two species.
Due to quorum-sensing, the self-propulsion velocity $v_S$ in Eq.~\eqref{eq:micro-model} depends on the
coarse-grained density fields $\tilde{\rho}_S(\mathbf{r},t)$ measured over a finite interaction scale $R$ via a short-ranged kernel $\hat{w}(r)$ \colrev{whose expression is given in Appendix.
The linear stability analysis performed below reveals that,
approximating $\tilde{\rho}_S(\mathbf{r},t) \simeq \rho_S(\mathbf{r},t)$ with $\rho_S(\mathbf{r},t) \equiv \sum_i \delta (\mathbf{r}-\mathbf{r}_{i,S}(t))$,}
the emergence of patterns from the dynamics described by~\eqref{eq:micro-model} is controlled by the dimensionless couplings
$\eta_{SS'}(\rho_A, \rho_B)\equiv \rho_S \partial \ln(v_S)/\partial \rho_{S'}$.
For $\eta_{SS'}<0$, a particle from species $S$ moves slower in higher $S'$ density regions, leading to an effective attraction via motility inhibition.
In turn, ${\eta}_{SS'}>0$ makes $S$ particles spend less time in regions of high $S'$ density, such that motility activation amounts to an effective repulsion.
The sign of $\eta_{SS}$ thus determines whether the effective interaction between particles of same species $S$ is attractive (repulsive), as a result of the self-inhibition (-activation) of their motilities.
For multiple species, nonreciprocity arises whenever $\rho_{S'}\eta_{S S'} \ne \rho_S\eta_{S' S}$, 
such that the dynamics~\eqref{eq:micro-model} cannot be coarse-grained to an effective equilibrium field theory~\cite{dinelli2022self}.
In particular, effective chasing interactions between species $A$ and $B$ are achieved when ${\eta}_{AB}{\eta}_{BA} < 0$.

For simplicity, we consider $v_S(\rho_A,\rho_B) = {v}_{0} L_{SA}(\rho_A) L_{SB}(\rho_B)$, where $L_{SS'}(x) > 0$ is a logistic function
such that $v_S$ varies monotonously with $\rho_A$ and $\rho_B$.
The values of the couplings $\eta_{SS'}$ can then be varied changing either the values of the densities $\rho_{A,B}$,
or the shape of the curves $L_{SS'}$. 
For convenience, we define a reference value $\rho_0$ such that $v_S(\rho_0,\rho_0) = v_0$ while $\eta_{SS'}^0 \equiv \eta_{SS'}(\rho_0,\rho_0)$
is the value at which $|L'_{SS'}|$ reaches its maximum.
Rescaling space and time, we set $R=1$ and ${v}_{0}=1$ so that the remaining control parameters of the 
microscopic model are the mean particle densities $\{\bar{\rho}_{S}\}$, the nominal couplings $\{\eta_{SS'}^0\}$, and $D_r$.
We moreover define the P\'eclet number $\mathrm{Pe}\equiv v_B(\bar{\rho}_A,\bar{\rho}_B)/D_r$ as a measure of the self-propulsion strength,
while the parameter $\chi \equiv \eta_{AB}(\bar{\rho}_A,\bar{\rho}_B)\eta_{BA}(\bar{\rho}_A,\bar{\rho}_B)$ is used to quantify nonreciprocity.
All simulations are performed in periodic domains of size $L_x \times L_y$, with total particle number ranging from $N = 10^4$ to $10^7$.
\colrev{Additional details about the microscopic model are given in Appendix.}

{\it Traveling patterns induced by chasing interactions.---}
To start with, we fix the mean particle densities $\bar{\rho}_{S} = \rho_0$,
such that the couplings $\eta_{SS'}$ evaluated at $\bar{\rho}_{A,B}$ are given by $\eta^0_{SS'}$ and ${\rm Pe} = v_0/D_r$.
For ${\rm Pe} \ll 1$ or $|\chi| \ll 1$, our observations are in line with predictions from the NRCH model~\cite{saha2020scalar,you2020nonreciprocity,frohoff2021suppression}.
Namely, if either of $\eta_{AA}^0$ or $\eta_{BB}^0$ are sufficiently negative, systems initialized in the homogeneous state are unstable. 
For $\chi > 0$, this instability leads to the phase separation of one of the two species, or demixing.
For negative $\chi$ and small systems, on the other hand, nonreciprocity gives rise to the formation of system spanning traveling bands rich in 
either of the two species and chasing each other (Fig.~\ref{FIG1}(a) and Supplemental Movie (SMov)~1).
As this TRS broken phase mainly relies on the presence of two phase-separated domains whose cohesion is maintained by self-inhibition,
TRS can be restored when one or both species see their self-coupling vanish, as shown in Fig.~\ref{FIG1}(b).

\begin{figure}[b!]
	\includegraphics[width=\columnwidth]{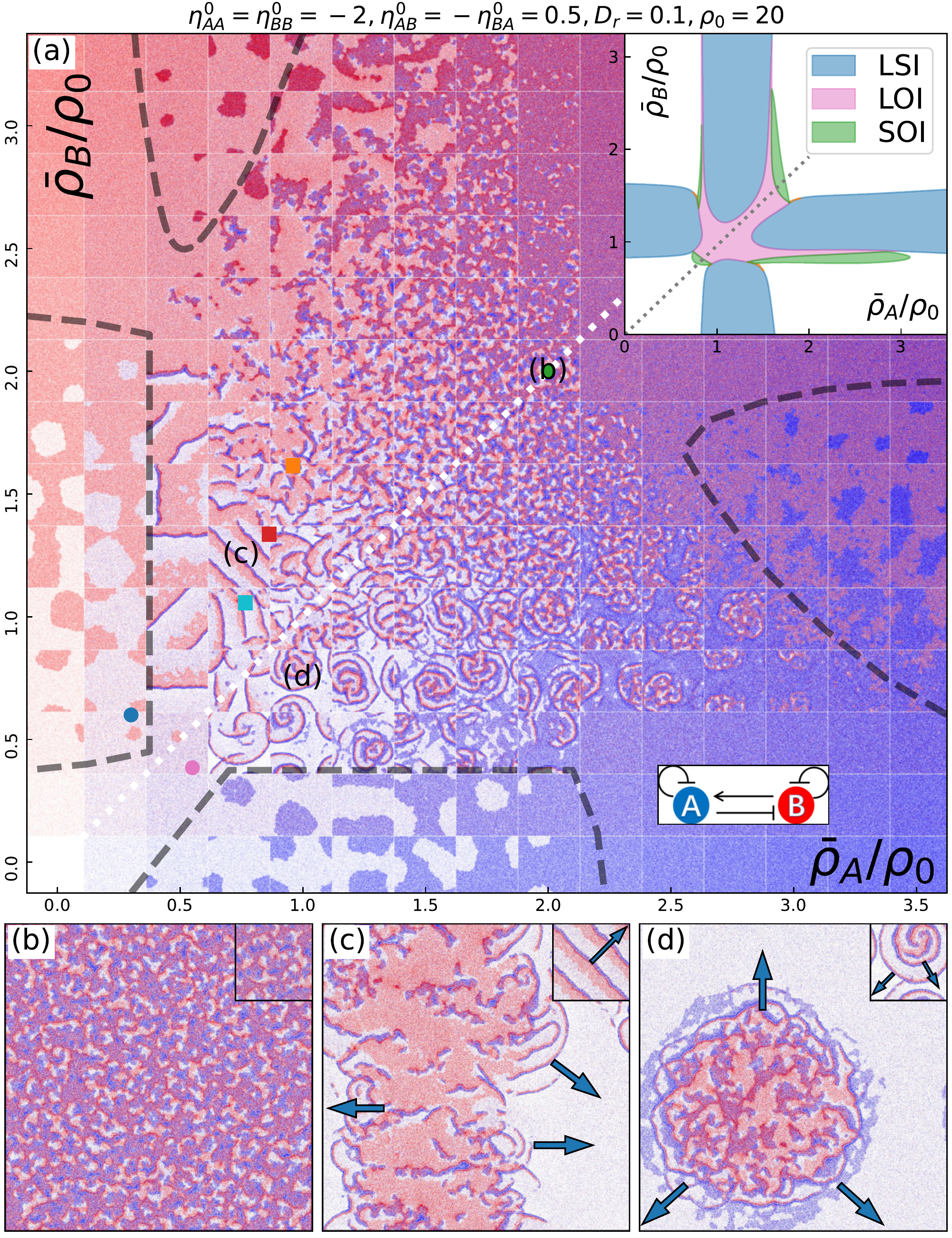}
	\caption{(a) Snapshots from particle-based simulations
	in the composition plane in square domains of size $L_x=40$.
	       The regions of static phase separation are indicated by the dashed lines as guides to the eye.
	       \colrev{Solid squares and dots respectively mark the compositions used to analyze the profiles of Figs.~\ref{FIG3}(a-c) and coarsening in Fig.~\ref{FIG3}(e).}
	       Inset: the corresponding linear stability diagram.
	       (b--d) Simulations in square domains of size $L_x = 160$, each with the same composition as its top right inset where $L_x = 40$.
	 }
	\label{FIG2}
\end{figure}

Strikingly, for $\chi < 0$
and $\rm Pe$ and $|\chi|$ large enough 
we find traveling patterns arising even at weak or vanishing self-attraction.
These patterns take the form of thin chasing bands which for small $|\chi|$ and moderate system size self-organize 
into a superposition of regular arrangements traveling along different directions (Fig.~\ref{FIG1}(c) and SMov~1).
Such structure is however rather fragile such that increasing $|\chi|$ (Fig.~\ref{FIG1}(d)), or in presence of moderate self-inhibition of motilities, 
it is destabilized and replaced by a chaotic chasing bands (CCB) phase (Fig.~\ref{FIG2}(b)).

Contrary to the regime of strong motility self-inhibition, the cohesion of the chasing bands here is ensured by the combined effects of $\rm Pe$ and $\chi < 0$.
For weak or vanishing self-attraction, traveling bands can only exist if particles inside them move coherently over large distances, as evidenced by the local polarization shown in the bottom row of Figs.~\ref{FIG1}(c,d).
\colrev{This feature, which is absent in strongly self-attracting mixtures (bottom row of Figs.~\ref{FIG1}(a,b)), remarkably arises despite the absence of explicit aligning interactions between particle velocities (Eq.~\eqref{eq:micro-model}).}
The residual particle flux to the outside of the bands induced by rotational noise is then balanced by the effect of nonreciprocal inter-species couplings.
To understand this, let us consider the case of Fig.~\ref{FIG1}(c,d) where species $A$ inhibits the motility of $B$ particles and is activated by them.
A cluster of $A$ particles traveling in the gas locally slows down $B$ particles, so that they aggregate at its rear.
In the case of vanishing self-attraction, $B$ particles can then follow the $A$ cluster only when they move in the same direction.
Similarly, the motility activation of $A$ particles provoked by a $B$ cluster leads them to reside at its front so long as they move in the same direction.
This emergent noise rectification mechanism relying on chasing interactions  allows the $AB$ cluster pair to continuously recruit particles from the gas.
To confirm this picture, we define the system- and time-averaged speed variance ${\rm Var}(v) \equiv \langle \langle |\dot{\bf r}_i|^2 \rangle_{i} - \langle|\dot{\bf r}_i|\rangle_{i}^2 \rangle_t$
which is nonzero only in the presence of motility induced patterns.
Scanning the $(-\chi,{\rm Pe})$ plane at $\eta_{AA}^0 = \eta_{BB}^0 = 0$, 
Fig.~\ref{FIG1}(e) shows that the CCB phase --- whose onset is characterized by an abrupt growth of ${\rm Var}(v)$ --- is found when both self-propulsion and nonreciprocity are sufficiently strong.

{\it Mesoscopic description \& linear stability analysis.---}
To get a theoretical understanding of the onset of CCB, we derived the coarse-grained description of the microscopic model.
\colrev{The full derivation uses standard coarse-graining techniques and is detailed in the Appendix. It leads to a pair of equations for the noise-averaged particle density and polarity.}
\colrev{Linearizing these equations around their solution with homogeneous densities $(\bar{\rho}_A,\bar{\rho}_B)$ and vanishing polarities, we find that its stability is determined by five parameters:}
$\eta_{AA}$,
$\eta_{BB}$,
$\chi$,
$\mathrm{Pe}$
and $\sigma_v\equiv {v}_A/{v}_B$, where $v_S$ and $\eta_{SS'}$ are evaluated at $(\bar{\rho}_A,\bar{\rho}_B)$.
Taking $\sigma_v = 1$ (see~\cite{supplement} for a discussion of the general case leading to similar results),
the eigenvalue controlling the stability is
$\lambda(q) = -\tfrac{q}{2}\gamma + \tfrac{ q}{2}\sqrt{\gamma^2-2\mu+2\sqrt{\alpha}}$,
where $q$ denotes the wavenumber of the perturbation while
$\gamma(q)\equiv (\mathrm{Pe} \, q)^{-1} + \mathrm{Pe}\,q/16$, $\mu\equiv 1 + \tfrac{1}{2}({\eta}_{AA}+{\eta}_{BB})$, and $\alpha \equiv \chi + \tfrac{1}{4}({\eta}_{AA}-{\eta}_{BB})^2$.

For $\alpha>0$, $\lambda$ is real as $q\to 0$ and a long-wave stationary instability (LSI) arises when $\mu<\sqrt{\alpha}$, i.e.\ $\mu <0$ or $(1 + {\eta}_{AA})(1 + {\eta}_{BB})<\chi$.
On the other hand, $\alpha = 0$ corresponds to an exceptional point~\cite{saha2020scalar} such that for $\alpha < 0$, the imaginary part of $\lambda(q)$ is always nonzero,
while its real part $\Re(\lambda)$ becomes positive iff~\cite{supplement}
\begin{equation} \label{eq:pos_lambda_re}
    2\mu \gamma^2(q)+\alpha = 
    2\mu \left(
    \frac{1}{\mathrm{Pe} \, q} + \frac{\mathrm{Pe}\,q}{16}
    \right)^2+\alpha
    <0.
\end{equation}
For $\alpha<0$ and $\mu<0$ the condition \eqref{eq:pos_lambda_re} is always true down to $q=0$, giving rise to a long-wave oscillatory instability (LOI).
The emergence of LSI and LOI are thus both mainly controlled by the self-interaction couplings, 
such that these instabilities may arise at arbitrary small $|\chi|$ and their range of existence is insensitive to $\mathrm{Pe}$.

Equation~\eqref{eq:pos_lambda_re} moreover shows that for $\alpha<0$ and $\mu>0$, although $\Re(\lambda)$ is negative as $q\to 0$, it may turn positive at finite $q$ when $\gamma^2(q) < -\alpha/2\mu$. As such a scenario can only happen when $\lambda$ is complex, it is associated to a short-wave oscillatory instability (SOI).
This is in contrast with the NRCH model~\cite{saha2020scalar,you2020nonreciprocity,frohoff2021suppression} or reaction-diffusion systems~\footnote{SOI is forbidden in standard binary reaction-diffusion systems~\cite{diego2018key}, but can occur under certain conditions, such as systems on curved surfaces~\cite{nishide2022pattern} or in the presence of differential flow~\cite{rovinsky1992chemical}} for which 
SOI is absent in binary mixtures.
In the limit $\eta_{AA} = \eta_{BB} = 0$ of vanishing self-couplings, the condition for SOI reduces to $\chi<-\tfrac{1}{2}$ (dotted line in Fig.~\ref{FIG1}(e)), while the lowest unstable wavenumber is given by ${\rm Pe^2}q_c^2 = -32(1+4\chi + 8\sqrt{(1/2+\chi)\chi})$.
Using the interaction range $R = 1$ as a natural lower cutoff for the admissible values of $q_c^{-1}$, we thus get a line in the $(\chi,{\rm Pe})$ plane that defines the boundary of the instability region shown in Fig.~\ref{FIG1}(e),
and which qualitatively agrees with the numerical simulations of the microscopic model.
We furthermore summarize the linear stability results for $\chi<0$ in the diagram of Fig.~\ref{FIG1}(f) drawn in the $(\eta_{AA},\eta_{BB})$ plane,
which confirms that SOI typically arises when both species present weak or vanishing self-couplings, while its range of existence increases with ${\rm Pe}$.

{\it CCB-gas phase coexistence.---}
As we now show, the range of existence of CCB actually extends beyond that of SOI, 
since it may correspond to one of the phases involved in the phase-separation configurations following LOI.
We now fix $\eta^0_{AA} = \eta^0_{BB} = -2$,  $\eta^0_{AB} = -\eta^0_{BA} = 0.5$, $D_r = 0.1$, 
and perform simulations of the microscopic model~\eqref{eq:micro-model} scanning the composition plane $(\bar{\rho}_A,\bar{\rho}_B)$.

The resulting phase diagram shown in Fig.~\ref{FIG2}(a) exhibits four distinctive lobes encircled by black dashed lines where static phase separation takes place~\cite{saha2020scalar}, 
in qualitative agreement with the location of LSI in the linear stability diagram (\colrev{blue regions in the inset of Fig.~\ref{FIG2}(a)}).
\colrev{Along the diagonal $\bar{\rho}_A=\bar{\rho}_B$ marked by a white dashed line in Fig.~\ref{FIG2}(a)}, the homogeneous phase becomes unstable below densities $\approx 2\rho_0$ and is superseded by 
the CCB phase (Fig.~\ref{FIG2}(b) and SMov~2) previously described.
Further decreasing the densities,
we observe system spanning traveling bands for $\bar{\rho}_B \gtrsim \bar{\rho}_A$ and rotating spirals for $\bar{\rho}_B \lesssim \bar{\rho}_A$.
\colrev{Examples of these patterns are shown in the insets of Figs.~\ref{FIG2}(c,d) and SMov~2.}
Increasing the system size, the CCB phase remains qualitatively unchanged, \colrev{as can be seen comparing the main panel and inset of Fig.~\ref{FIG2}(b).}
On the other hand, both traveling bands and spirals found at lower densities are replaced by a macroscopic CCB domain coexisting with a dilute homogeneous gas (Fig.~\ref{FIG2}(c,d)). 
\colrev{Accordingly, the locations of phase-separated CCB domains qualitatively correspond to regions of the composition plane where LOI is found (maked in magenta in the inset of Fig.~\ref{FIG2}(a)).
The system-wide CCB phase, on the other hand, borders the uniform high-density phase, in line with the location of SOI highlighted in green in the inset of Fig.~\ref{FIG2}(a).}

\begin{figure}
	\includegraphics[width=\columnwidth]{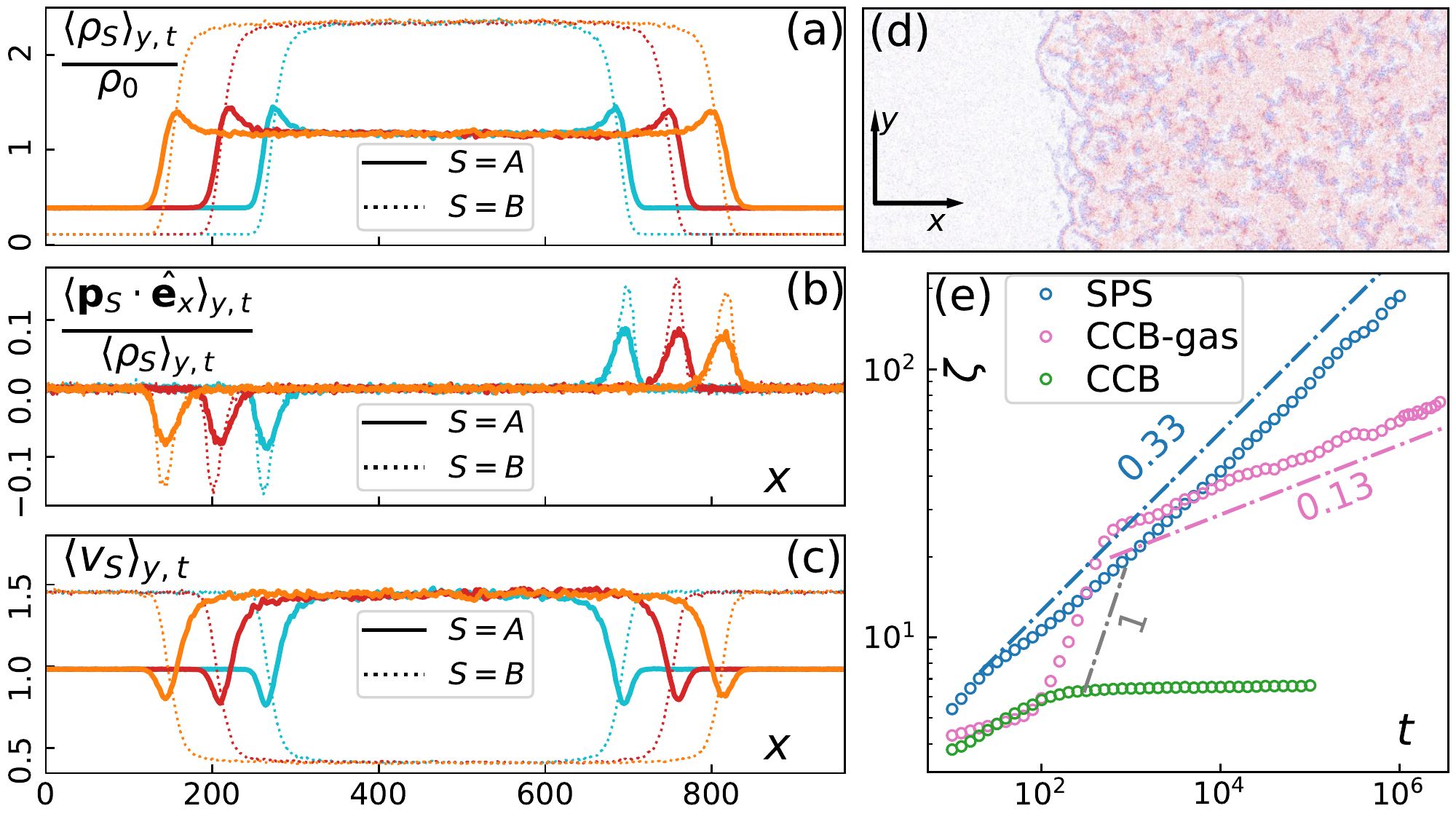}
	\caption{(a--c)  $y$- and time-averaged density (a), $x$-component of the polarity (b) and speed (c) profiles for both species at various compositions along the tie line with $L_x=6L_y=960$.
		     (d) CCB-gas interface for the configuration corresponding to the red curves in (a--c).
	         (e) Typical domain size $\zeta$ evaluated from the first moment of the structure factor~\cite{supplement} as function of time for randomly initialized systems in the static phase separation (SPS), 
	         CCB-gas phase coexistence and pure CCB regimes.
	         For each curve in (a--c,e), the corresponding composition is marked by a symbol of the same color in Fig.~\ref{FIG2}(a).
	 }
	\label{FIG3}
\end{figure}

To further characterize the CCB-gas coexistence phase, we performed simulations within this regime in a large rectangular domain.
This way, the CCB domain connects with itself along the shortest dimension $y$, leading to a well-defined interface (Fig.~\ref{FIG3}(d)).
\colrev{The $y$ and time-averaged density profiles shown in Fig.~\ref{FIG3}(a) indeed highlight two bulk phases of nearly constant densities.
The values taken by $\langle\rho_{A,B}\rangle_{y,t}$ far away from the interfaces thus define a pair of points in the composition plane which can be joined by a tie line.
Consistently with the usual picture of phase separation, shifting $(\bar{\rho}_A,\bar{\rho}_B)$ along this tie line 
changes the relative proportions of the gas and CCB phases while the corresponding bulk densities are left unchanged.
}

TRS is however obviously broken at mesoscopic scales in the CCB-gas coexistence phase,
since chasing bands are constantly created in the dense CCB domain and expelled in the gas, where they quickly dissolve (Fig.~\ref{FIG3}(d) and SMov~3).
The outward mass flux from the chasing bands must then be balanced by the diffusive flux from the resulting excess particles in the gas, thereby maintaining the cohesion of the CCB domain.
The nontrivial structure of the CCB-gas interface is also highlighted by the presence of local polarization pointing towards the dilute regions (Fig.~\ref{FIG3}(b)), at odds with the usual MIPS phenomenology~\cite{Omar2020PRE,Mahault2022}.
We moreover note from Figs.~\ref{FIG3}(a,c) that $A$ particles accumulate at interfaces where their self-propulsion speed is lowest, 
but move on average faster in the dense CCB domain than in the dilute gas.

Remarkably, the nonequilibrium nature of the CCB-gas phase coexistence also emerges over macroscopic scales,
as revealed by the anomalous coarsening behavior shown in  Fig.~\ref{FIG3}(f).
Although in the regimes of pure CCB and static phase separation the coarsening is arrested or follows the Lifshitz–Slyozov $t^{1/3}$ law~\cite{Lifshitz1961JPCS} (green and blue symbols in Fig.~\ref{FIG3}(f)),
the late-time coarsening of phase-separated CCB domains
is characterized by an exponent $\approx 0.13$ (magenta symbols in Fig.~\ref{FIG3}(f)),
significantly smaller than the $\tfrac{1}{3}$ value expected in passive systems~\cite{BrayAdvPhys2002} and for the coarsening of dense MIPS domains~\cite{Stenhammar2014SoftMatt,shi2020self,caporusso2020motility,caporusso2022dynamics,Note2}.
We rationalize this result by noting that larger CCB domains generate more bands ---and thus expel more mass into the surrounding gas--- than smaller domains, which may naturally slow down coarsening.
\phantom{\footnote{We note that for an underdamped dynamics the mapping of MIPS to an equilibrium coarse-grained model at large scales may not hold~\cite{mandal2019motility}.}}

Using a minimal model including self-propulsion and nonreciprocity, we have shown how the combination of these two sources of activity leads to a chaotic chasing band phase.
Since it is involved in phase-separated configurations at large nonreciprocity, this phase is moreover found in a large portion of the phase diagram.
Additionally, although the corresponding phase-separated domains are globally static, they still defy an equilibrium mapping as evidenced by the observed abnormal coarsening behavior.
\colrev{As they allow for the design of programmable quorum-sensing motility responses, both genetically engineered {\it E. coli}~\cite{Weiss2008} and light-controlled microswimmers~\cite{Bregulla2014ACS,Gomez-Solano2017,Massana2022NatCom,yang2021controlling} offer promising experimental platforms to observe this new type of self-organized behavior.
}


\acknowledgments
We thank Xiaqing Shi and Hugues Chat\'e for their critical reading of our manuscript.
This work has received support from the Max Planck School Matter to Life and the MaxSynBio Consortium, which are jointly funded by the Federal Ministry of Education and Research (BMBF) of Germany, and the Max Planck Society.
\\
\appendix

\setcounter{figure}{0}
\renewcommand\thefigure{A\arabic{figure}}
\setcounter{equation}{0}
\renewcommand\theequation{A\arabic{equation}}

\colrev{
\noindent{\it Appendix on simulations of the microscopic model.---}
Here, we give additional details on the microscopic model~\eqref{eq:micro-model} and discuss the underlying assumptions behind its formulation.
The Langevin equations~\eqref{eq:micro-model} were integrated using the Euler-Maruyama scheme with time resolution $\Delta t = 0.1$.

The local density fields $\tilde{\rho}_{S}({\bf r},t)$ entering the expression of the quorum-sensing interaction in Eq.~\eqref{eq:micro-model} were computed as
 $
 \tilde{\rho}_S(\mathbf{r},t) = \sum_i \hat{w}(|\mathbf{r}-\mathbf{r}_{i,S}(t)|),
 $
 with the linear weight function
 \begin{equation*} \label{seq:weight-func}
 \hat{w}(r) = \begin{cases}
    \frac{3}{\pi R^2}(1-\frac{r}{R}) \quad &\mathrm{if} \quad r < R,\\
    0 \quad &\mathrm{if} \quad r\geq R,
 \end{cases}
 \end{equation*}
 that satisfies the normalization condition $\int_{\mathbb{R}^2} {\rm d}\mathbf{r}\, \hat{w}(|\mathbf{r}|)=1$.
 As mentioned in the main text, the dependency of the particles self-propulsion speed with their density is modeled as $v_S(\tilde{\rho}_A,\tilde{\rho}_B) = {v}_{0} L_{SA}(\tilde{\rho}_A) L_{SB}(\tilde{\rho}_B)$ with
\begin{equation*}
    L_{SS'}(x) \equiv 1 + \kappa \tanh \left(\frac{\eta_{SS'}^0}{\kappa} \frac{x-\rho_0}{\rho_0}\right).
\end{equation*}
Therefore, the reference density value $\rho_0$ satisfies $L_{SS'}(\rho_0)=1$ and $\partial_x L_{SS'}(\rho_0)=\eta_{SS'}^0$. 
Here, $\kappa$ controls the lower and upper limits of $L_{SS'}(x)$.
To prevent jamming of particles at high densities, we set  $\kappa=0.7$ so that the minimum speed of species $S$ is given by $\approx 0.09\,v_{0}$.
 
 To keep the microscopic model minimal, we have assumed a direct dependency of the particle motility in the local density fields.
 We expect this approximation to be well-verified for synthetic colloids controlled by optical feedback loops, which allow for a fast motility response to dynamical variations of the local density~\cite{Gomez-Solano2017,bauerle2018self}.
 In the case of microorganisms communicating via chemical signals, the limit of instantaneous quorum-sensing response requires that the timescales associated with diffusion of chemicals and the internal gene regulatory network are fast as compared to the spatial dynamics of particles. 
 Whether this assumption is verified depends on the system of interest, but the limit of fast quorum-sensing response should in general qualitatively capture the relevant dynamics~\cite{curatolo2020cooperative}.

Another simplification used in the microscopic model~\eqref{eq:micro-model} consists in neglecting the short-range repulsive interactions between particles.
This assumption is well-justified so long as the range of quorum-sensing interactions $R$ is much larger than the typical particles size $\sigma$.
In synthetic systems, $R$ can be tuned arbitrarily, while $R = 10\sigma$ was used in Ref.~\cite{bauerle2018self}.
In the case of quorum-sensing regulated by chemicals, the scale $R$ is set by the typical distance a signaling molecule can diffuse before being degraded. 
Estimates for the acyl-homoserine lactone molecules that mediate the quorum-sensing interactions of bacteria in Refs.~\cite{Liu2011Science,curatolo2020cooperative} can reach a few millimeters~\cite{marenda2016modeling}, thus several orders of magnitude larger then the typical size of a bacterium. \\}

\colrev{
\noindent {\it Appendix on the coarse-grained equations.---}
To derive the field theory describing the binary mixture of self-propelled particles with quorum-sensing interactions, we follow the coarse-graining framework developed for a single-species~\cite{10.1088/1361-648X/acc440}.
As a starting point, we consider the many-body
probability distribution $P(\mathbf{X}, t)$ with $\mathbf{X} \equiv \{\mathbf{r}_1, \mathbf{r}_2,\dots,\mathbf{r}_N, \theta_1, \theta_2, \dots,\
\theta_N\}$,
and where $N$ denotes the total number of particles of the two species.
From the Langevin formulation~\eqref{eq:micro-model}, we get the statistically equivalent Fokker-Planck equation
\begin{equation} \label{seq:P_X_t}
    \partial_t P = -\sum_{i=1}^N \left[
        \nabla_{\mathbf{r}_i}\cdot \left(v_{s_i}\hat{\mathbf{u}}(\theta_i) P\right)
        - 
        D_r \partial_{\theta_i}^2P\right],
\end{equation}
where $\hat{\mathbf{u}}(\theta)=(\cos \theta, \sin \theta)$ and $s_i = A$ or $B$ denotes the species of particle $i$.
Without loss of generality, we set $s_i=A$ for $i=1, 2,\dots, N_A$ and $s_i=B$ for $i=N_A+1, N_A+2, \dots,N$ with $N_A$ the total number of $A$ particles.
The one-particle probability densities $f_S(\mathbf{r},\theta, t)$ are obtained from $P$ by integrating over all degrees of freedom except that of one particle of species $S$.
Hence, we have $f_A(\mathbf{r}_1, \theta_1, t) \equiv N_A\left(\prod_{i=2}^N \int \mathrm{d}\mathbf{r}_i\int_0^{2\pi}\mathrm{d}\theta_i\right)
P(\mathbf{X}, t)$
while $f_B$ is defined in a similar way.
Integrating Eq.~\eqref{seq:P_X_t} over the relevant coordinates, we thus determine the dynamics of the $A$ and $B$ single-particle distributions.
For $f_{A}$ (the generalization to the species $B$ being straightforward), we obtain
\begin{align} \label{seq:fA}
    \partial_t f_A(\mathbf{r}_1, \theta_1, t) & = - \nabla_{\bf{r}_1}\cdot\left[{\cal G}_A(\mathbf{r}_1, \theta_1, t) \hat{\bf{u}}(\theta_1) f_A(\mathbf{r}_1, \theta_1, t)\right] \nonumber \\
    &  + D_r \partial_{\theta_1}^2 f_A(\mathbf{r}_1, \theta_1, t),
\end{align}
where ${\cal G}_A(\mathbf{r}_1, \theta_1, t) \equiv 
    \left(\prod_{k=2}^N \int \mathrm{d}\mathbf{r}_k\right)v_{A}({\bf X}_r) g_A({\bf X}_r,\theta_1,t)$,
and we have used the shorthand notation $\mathbf{X}_r \equiv \{\mathbf{r}_1,\dots,\mathbf{r}_N\}$.
The function $g_A$ is defined by $N_A \prod_{k=2}^N\int_0^{2\pi}\mathrm{d}\theta_k P({\bf X},t) \equiv f_A(\mathbf{r}_1, \theta_1, t) g_A({\bf X}_r,\theta_1,t)$.
In general, the quorum-sensing interaction term ${\cal G}_A$ in~\eqref{seq:fA} depends on the full many-body distribution.
However, we now use the fact that the particle speed is formally a function of $\tilde{\rho}_{A,B}({\bf X}_r)$, which in the mean field limit can be expressed in terms of the single particle densities: 
\begin{align*}
    \tilde{\rho}_S({\bf X}_r) & =
    \sum_j\hat{w}(|\mathbf{r}_1 - \mathbf{r}_j(t)|)\delta_{s_j, S} \\
    & \simeq \int \mathrm{d}\mathbf{r}'\hat{w}(|\mathbf{r}_1 - \mathbf{r}'|)\phi_S(\mathbf{r}',t) \equiv \tilde{\phi}_S({\bf r}_1,t),
\end{align*}
where the $S$ particle density is formally defined as $\phi_S({\bf r},t) \equiv \int_0^{2\pi}{\rm d}\theta f_S(\mathbf{r}, \theta, t)$.
We expect this approximation to be reasonably valid in sufficiently dense systems.
In the mean field limit considered here, the function $v_A$ in the expression of ${\cal G}_A$ thus only depends on the position variable ${\bf r}_1$.
Using that by definition $\left(\prod_{k=2}^N \int \mathrm{d}\mathbf{r}_k\right) g_A({\bf X}_r,\theta_1,t) = 1$,
we therefore obtain after dropping indices on the $\bf r$ and $\theta$ variables:
\begin{equation} \label{seq:fS}
    \partial_t f_S = - \nabla\cdot\left[v_{S}(\tilde{\phi}_A,\tilde{\phi}_B) \hat{\bf{u}}(\theta) f_S \right] + D_r \partial_{\theta}^2 f_S.
\end{equation}
To simplify Eq.~\eqref{seq:fS} further,
we use the local approximation $\tilde{\phi}_S(\mathbf{r})\approx\phi_S(\mathbf{r})$, such that $v_S(\tilde{\phi}_A,\tilde{\phi}_B)\approx v_S(\phi_A,\phi_B)$.

Due to angular diffusion, we expect the dynamics of the system to be well captured over long timescales by that of the low order orientational moments of the distributions $f_A$ and $f_B$.
Therefore, we expand the distribution $f_S$ in angular Fourier modes: $f_S(\mathbf{r},\theta,t)=(2\pi)^{-1}\sum_{k=-\infty}^{\infty} f_{k, S}(\mathbf{r},t) \exp(-ik\theta)$.
It is straightforward to check that the first three modes of $f_S$ correspond to the complex representation of the density $\phi_S({\bf r},t)$, polarity $\mathbf{p}_S(\mathbf{r},t)$ and nematic order $\mathbf{Q}_S(\mathbf{r},t)$ fields.
For the derivation below, it is convenient to work with complex notations for which $\hat{\bf u}(\theta) \leftrightarrow e^{i\theta}$. 
We therefore define the complex gradient $\triangledown \equiv \partial_x + i \partial_y$, and obtain from Eq.~\eqref{seq:fS}
\begin{align} \label{eq:hierarchies}
	\partial_t f_{k,S} & =
	-\frac{1}{2} \triangledown^* \left[v_S(\phi_A,\phi_B) f_{k+1,S}\right] \nonumber \\
	& - \frac{1}{2} \triangledown\left[v_S(\phi_A,\phi_B) f_{k-1,S}\right]
	- D_r k^2 f_{k,S},
\end{align}
where star denotes complex conjugate.
The equation for the $k^{\rm th}$ angular mode of $f_S$ contains a linear damping term $- D_r k^2 f_{k,S}$. 
Considering the long time and large scale limits, we thus enslave the dynamics of the high order modes to that of the slow ones.
As the densities are the only conserved fields, it is customary in this context to retain only them as hydrodynamic fields, while enslaving higher order modes~\cite{cates2015motility}.
Motivated by the presence of local polarization in the CCB phase, here we instead retain both $\phi_S$ and the polarity fields
$\mathbf{p}_S(\mathbf{r},t)$.
Namely, neglecting $f_{k,S}$ for $k \ge 3$, 
we get from~\eqref{eq:hierarchies} 
closed equations for $\phi_S$, $f_{1,S}$ and $f_{2,S}$.
Setting $\partial_t f_{2,S} = 0$, we solve the equation for $f_{2,S}$ which leads to $f_{2,S}=-(8D_r)^{-1}\triangledown(v_S f_{1,S})$.
Replacing this expression in the equation for $f_{1,S}$, 
we get after going back to vector notations
\begin{subequations} \label{eq:rho-p}
\begin{align}
		\partial_t \phi_S = & -\nabla \cdot (v_S \mathbf{p}_S)\label{eq:rho-p-a},\\
		\partial_t \mathbf{p}_S = & 
		- \frac{1}{2} \nabla (v_S \phi_S) - D_r \mathbf{p}_S 
		+ \frac{v_S}{16D_r}\Delta \left(v_S\mathbf{p}_S\right) \nonumber \\
		& + (8D_r)^{-1}\left[\nabla(v_S \mathbf{p}_S)\right]_{\rm ST} \cdot \nabla v_S,
		\label{eq:rho-p-b}
\end{align}
\end{subequations}
where $\left[{\bf A}\right]_{\rm ST} \equiv \tfrac{1}{2}\left[{\bf A} + {\bf A}^{T} - {\bf I}{\rm Tr}({\bf A})\right]$ is the symmetric and traceless part of the tensor $\bf A$.
Performing the linear stability analysis of Eqs.~\eqref{eq:rho-p} around their homogeneous disordered solution $\phi_{S} = \bar{\rho}_S$ and ${\bf p}_S = {\bf 0}$, we obtain the results presented in the main text.
}
 
\bibliography{NRQS_main.bib}

\end{document}


\title{Supplemental Material to ``Dynamical pattern formation without self-attraction in quorum-sensing active matter: the interplay between nonreciprocity and motility"}

\author{Yu Duan}
\affiliation{Max Planck Institute for Dynamics and Self-Organization (MPI-DS), 37077 G\"ottingen, Germany}

\author{Jaime Agudo-Canalejo}
\affiliation{Max Planck Institute for Dynamics and Self-Organization (MPI-DS), 37077 G\"ottingen, Germany}

\author{Ramin Golestanian} \email{ramin.golestanian@ds.mpg.de}
\affiliation{Max Planck Institute for Dynamics and Self-Organization (MPI-DS), 37077 G\"ottingen, Germany}
\affiliation{Rudolf Peierls Centre for Theoretical Physics, University of Oxford, Oxford OX1 3PU,
United Kingdom}

\author{Beno\^{\i}t Mahault} \email{benoit.mahault@ds.mpg.de}
\affiliation{Max Planck Institute for Dynamics and Self-Organization (MPI-DS), 37077 G\"ottingen, Germany}

\maketitle

\tableofcontents

\section{Linear instability analysis of the homogeneous disordered solution}

For clarity, we recall the hydrodynamic equations obtained in the main text:
\begin{subequations} \label{seq:rho-p}
	\begin{align}
		\partial_t \phi_S &= -\nabla \cdot (v_S \mathbf{p}_S),\label{seq:rho-p-a}\\
		\partial_t \mathbf{p}_S &= \frac{v_S}{16D_r}\Delta \left(v_S\mathbf{p}_S\right) - \frac{1}{2} \nabla (v_S \phi_S) - D_r \mathbf{p}_S + \frac{1}{8D_r}\left[\nabla(v_S \mathbf{p}_S)\right]_{\rm ST} \cdot \nabla v_S, \label{seq:rho-p-b}
	\end{align}
\end{subequations}
where $\left[{\bf A}\right]_{\rm ST} \equiv \tfrac{1}{2}\left[{\bf A} + {\bf A}^{T} - {\bf I}{\rm Tr}({\bf A})\right]$ is the symmetric and traceless part of the tensor $\bf A$.

As mentioned in the main text, the homogeneous disordered solution of Eqs.~\eqref{seq:rho-p} can undergo three types of instabilities: long-wave stationary instability (LSI), long-wave oscillatory instability (LOI) and short-wave oscillatory instability (SOI).
The criterion we choose to decide the nature of the instability is based on the minimal wavenumber $q_\mathrm{min}$ at which the real part of the growth rate $\lambda(q)$ becomes positive.
Namely, $q_{\rm min} > 0$ corresponds to a short-wave instability while $\Re(\lambda(q_{\rm min}\to 0)) > 0$ defines long-wave instabilities.
In turn, the instability is said to be stationary if $\Im(\lambda(q_\mathrm{min}))=0$ and oscillatory otherwise.
Note that because of interaction range $R=1$, we consider wavenumbers $q\in[0, 1]$, as smaller scales are not expected to be captured by the coarse-grained description of the microscopic model.

\subsection{Linearized equations} \label{sec:lin-eq-full-model}

We linearize Eqs.~\eqref{seq:rho-p} around the homogeneous disordered solution
$\{\phi_S(\mathbf{r})=\bar{\phi}_S,\mathbf{p}_S(\mathbf{r})={\bf 0}\}$, such that the mean field speed becomes $\bar{v}_S\equiv v_S(\bar{\phi}_A,\bar{\phi}_B)$. 
The dynamics of the Fourier transforms of perturbations $\{\delta\hat{\phi}_S(\mathbf{q},t), \hat{\mathbf{p}}_S(\mathbf{q},t)\}$ then read
\begin{subequations} \label{seq:rho-p-non-local-linear-Fourier}
\begin{align}
    \partial_t \delta\hat{\phi}_S &= -i\bar{v}_S \mathbf{q}\cdot \hat{\mathbf{p}}_S, \label{seq:rho-p-non-local-linear-Fourier-a}\\
    \partial_t \hat{\mathbf{p}}_S &= -\left(D_r + q^2\frac{\bar{v}_S^2}{16 D_r}\right)\hat{\mathbf{p}}_S - \frac{i \mathbf{q} \bar{v}_S}{2}\left(
	\delta \hat{\phi}_S + \eta_{SA} \delta \hat{\phi}_A + \eta_{SB}\delta\hat{\phi}_B
	\right), \label{seq:rho-p-non-local-linear-Fourier-b}
\end{align}
\end{subequations}
where $\eta_{SS'} \equiv \bar{\phi}_S \partial \ln(v_S)/\partial \phi_{S'}$ are dimensionless couplings evaluated at $(\bar{\phi}_A,\bar{\phi}_B)$.
Substituting $\hat{\mathbf{p}}_S = q^{-2}[\hat{p}_{\parallel,S} \mathbf{q}_\parallel + \hat{p}_{\perp,S} \mathbf{q}_\perp]$ into Eqs.~\eqref{seq:rho-p-non-local-linear-Fourier}, we find that the component of $\hat{\bf p}_S$ orthogonal to $\bf q$ is decoupled from the other perturbations and cannot generate any instability.
Moreover, as the equations for $\delta\hat{\phi}_S$ and $\hat{p}_{\parallel,S}$ only depend on $q$ we denote
$\delta\mathbf{u}(q,t) \equiv (\delta \hat{\phi}_A(q,t), \delta \hat{\phi}_B(q,t), \hat{p}_{\|,A}(q,t), \hat{p}_{\|,B}(q,t))$, which leads to the linear system
\begin{equation} \label{seq:u_q_t}
    \partial_t \delta \mathbf{u}(q, t) = q\bar{v}_B \underline{\mathbf{M}}(q) \delta \mathbf{u}(q, t),
\end{equation}
where
\begin{equation*} \label{seq:det-matrix}
    \underline{\mathbf{M}}(q) \equiv \begin{pmatrix}
        0 & 0 & -i\sigma_v & 0 \\
        0 & 0 & 0 & -i \\
        -i \sigma_v \omega_{AA} & -i\sigma_v\omega_{AB} & -\gamma_A(q) & 0 \\
        -i\omega_{BA} & -i\omega_{BB} & 0 & -\gamma_B(q)
    \end{pmatrix},
\end{equation*}
and where we have defined
\begin{equation*}
    \sigma_v  \equiv \frac{\bar{v}_A}{\bar{v}_B},\quad
    \mathrm{Pe} \equiv \frac{\bar{v}_B}{D_r},\quad
    \omega_{SS'} \equiv \frac{\delta_{S,S'}+{\eta}_{SS'}}{2},\quad
    \gamma_A(q)  \equiv \frac{1}{\mathrm{Pe}\, q} + \frac{\sigma_v^2\mathrm{Pe}\,q}{16},\quad
    \gamma_B(q)  \equiv \frac{1}{\mathrm{Pe}\, q} + \frac{\mathrm{Pe}\,q}{16}.
\end{equation*}
The growth rate of the linear system \eqref{seq:u_q_t} is $\lambda(q) = q \bar{v}_B \tilde{\lambda}(q)$, where $\tilde{\lambda}(q)$ are the eigenvalues of the matrix $\underline{\mathbf{M}}(q)$ which can be evaluated from the roots of the determinant
\begin{equation} \label{seq:M_minus_lambda}
    \begin{aligned}
     |\underline{\mathbf{M}}-\tilde{\lambda}\underline{\mathbf{I}}|
    &= \tilde{\lambda}^4 + (\gamma_A+\gamma_B)\tilde{\lambda}^3 + (\gamma_A\gamma_B + \omega_{BB}+\sigma_v^2\omega_{AA})\tilde{\lambda}^2
       + (\gamma_A\omega_{BB}+\sigma_v^2\gamma_B\omega_{AA})\tilde{\lambda} + \sigma_v^2 (\omega_{AA}\omega_{BB}- \omega_{AB}\omega_{BA}) ,
    \end{aligned}
\end{equation}
which here is a quartic polynomial of $\tilde{\lambda}$ with $q$-dependent coefficients.
Thus, setting the distance units such that $R = 1$, the value of $\tilde{\lambda}(q)$ at fixed $q$ is set by five dimensionless parameters:
$$\mathrm{Pe},\quad\sigma_v,\quad\chi_A\equiv {\eta}_{AA}+1=2\omega_{AA},\quad\chi_B\equiv{\eta}_{BB}+1=2\omega_{BB},\quad\chi\equiv{\eta}_{AB}{\eta}_{BA}=4\omega_{AB}\omega_{BA} .$$
Solving $|\underline{\mathbf{M}}-\tilde{\lambda}\underline{\mathbf{I}}|=0$ can always be done numerically, while obtaining a tractable analytical expression for the dispersion relation for the general case is difficult.
However, we show below that $\chi < 0$ (inter-species couplings with opposite signs) is a necessary condition for LOI and SOI to emerge, while only LSI takes place for $\chi\geq 0$. 

\subsection{Special case for $\sigma_v=1$}

For $\sigma_v=1$, we can get the analytical expression for the eigenvalue $\lambda$ with the largest real part:
\begin{equation} \label{eq:lambda_sigma_1}
    \lambda(q)=\frac{1}{2}\left(-\gamma + \sqrt{\gamma^2-2\mu+2\sqrt{\alpha}}\right)q,
\end{equation}
where
$$\gamma(q)\equiv\frac{1}{\mathrm{Pe} q}+\frac{\mathrm{Pe}q}{16}, \qquad \mu\equiv \frac{\chi_A+\chi_B}{2}, \qquad \alpha\equiv\chi + \frac{(\chi_A-\chi_B)^2}{4}.$$
As discussed in the main text, for $\alpha>0$, $\lambda$ is real as $q\to 0$, and is positive once $\mu<\sqrt{\alpha}$, i.e.\ $\mu <0$ or $\chi_A\chi_B<\chi$. We thus find the long-wave stationary instability (LSI) in this situation.

On the other hand, for $\alpha < 0$, $\lambda$ will always have nonzero imaginary part. 
Setting $a+bi=\sqrt{\gamma^2-2\mu+2\sqrt{\alpha}}$ such that $\lambda = \frac{1}{2}(-\gamma+a+bi)q$, where both $a$ and $b$ are real, we find that $a$ solves
$a^4 - (\gamma^2 -2\mu)a^2 + \alpha = 0$ while the largest solution reads
\begin{equation} \label{eq:a1}
    a = \sqrt{\frac{\gamma^2 - 2\mu + \sqrt{\Delta_a}}{2}},
\end{equation}
where $\Delta_a\equiv (\gamma^2 - 2 \mu)^2 - 4 \alpha> 0$ when $\alpha< 0$.
Since $\Re(\lambda)=\frac{1}{2}(-\gamma + a)$, using Eq.~\eqref{eq:a1} we conclude that a sufficient and necessary condition for $\Re(\lambda)>0$ is
\begin{equation} \label{eq:condi_pos_lambda_neg_alpha}
    2\mu\gamma^2 + \alpha < 0.
\end{equation}

For $\alpha<0$ and $\mu < 0$, the condition~\eqref{eq:condi_pos_lambda_neg_alpha} is true down to $q=0$, indicating the presence of a long-wave oscillatory instability (LOI).
For $\alpha<0$ and $\mu > 0$,  since $\gamma(q)\to +\infty$ as $q\to0$, the condition~\eqref{eq:condi_pos_lambda_neg_alpha} cannot hold as $q\to0$, leading to $\Re(\lambda) \le 0$ at vanishing $q$. However, $\Re(\lambda)$ can turn positive at finite $q$ when $\gamma^2 < -\alpha/2\mu$, giving rise to a short-wave oscillatory instability (SOI).
For the case of vanishing motility self-inhibition $\eta_{AA}^0=\eta_{BB}^0=0$, the condition~\eqref{eq:condi_pos_lambda_neg_alpha} is reduced to $2\gamma^2 + \chi<0$, which is further equivalent to
\begin{equation}  \label{eq:condi_pos_lambda_vanishing_self_inhibition}
    \frac{\mathrm{Pe}^4}{16^2} q^4 + \frac{(1+4\chi)\mathrm{Pe}^2}{8} q^2 + 1 < 0.
\end{equation}
The condition~\eqref{eq:condi_pos_lambda_vanishing_self_inhibition} can be verified only when 
$\chi(2\chi + 1)>0$, which for negative $\chi$ imposes $\chi < -\tfrac{1}{2}$.
Under this precondition, the lowest unstable wavenumber is given by
\begin{equation}
    q_c^2 = -\frac{32}{\mathrm{Pe}^2}\left(1+4\chi + 8\sqrt{\tfrac{\chi}{2}(1+2\chi)}\right).
\end{equation}
In practice, we use $q_{\mathrm{cut}}\simeq 1/R = 1$ as the cutoff wavenumber above which the coarse-grained description~\eqref{seq:rho-p} breaks down. 
Therefore, for given values of $\chi$ and $\mathrm{Pe}$, SOI takes place when the corresponding $q_c$ exists and is smaller than $q_{\mathrm{cut}}$.

\subsection{The general case}
For the general case with $\sigma_v\neq 1$ an analytical expression for $\lambda$ from the determinant \eqref{seq:M_minus_lambda} is not available.
Nevertheless, for long-wave instabilities we only need to consider the limit of vanishing wavenumber. On the other hand, the onset of short-wave instabilities can be detected using the Routh-Hurwitz stability criterion~\cite{gantmacher2005applications}, without resorting to evaluating the roots of the determinant \eqref{seq:M_minus_lambda}.
Here, we show these approaches lead to similar conclusions for the linear stability of the homogeneous disordered solution as those obtained in the previous section.

\subsubsection{Long-wave instabilities}\label{sec:LI}

To characterize  long-wave instabilities analytically, we only need to study the linear dispersion relation in the limit $q\to0$. 
As we expect that the unstable growth rate satisfies $\tilde{\lambda}(q) \sim q$ for $q \to 0$,
Eq.~\eqref{seq:M_minus_lambda} leads in this limit to
\begin{equation*}
    \frac{4\sigma\tilde{\lambda}^2}{{\rm Pe}^2 q^2}
       + (\chi_A + \sigma\chi_B)\frac{2\tilde{\lambda}}{{\rm Pe} q} 
       + \chi_A\chi_B- \chi = 0,
\end{equation*}
where $\sigma \equiv 1/\sigma_v^2$.
We therefore obtain
\begin{equation*}
    \lambda(q) \underset{q\to0}{\simeq} -\frac{q^2 {\rm Pe} \bar{v}_B}{4\sigma} \left[ \chi_A + \sigma\chi_B \pm \sqrt{ (\chi_A - \sigma\chi_B)^2 + 4\sigma\chi } \right]
    = -\frac{q^2 {\rm Pe} \bar{v}_B}{4} \sqrt{\frac{|\chi|}{\sigma}} \left[ \tilde{\chi}_A + \tilde{\chi}_B \pm \sqrt{ (\tilde{\chi}_A - \tilde{\chi}_B)^2 + 4{\rm sgn}(\chi) } \right],
\end{equation*}
where we have introduced the reduced parameters $\tilde{\chi}_A\equiv\chi_A/\sqrt{\sigma|\chi|}$ and $\tilde{\chi}_B\equiv\sqrt{\sigma/|\chi|}\chi_B$.
It follows that the onset of long-wave instability is dictated by only three parameters: $\mathrm{sgn}(\chi)$, $\tilde{\chi}_A$, and $\tilde{\chi}_B$. 
For $\chi > 0$, a long-wave instability arises once $\tilde{\chi}_A\tilde{\chi}_B < 1$ and is stationary. 
On the other hand, for negative $\chi$, the long-wave instability occurs when $\tilde{\chi}_A+\tilde{\chi}_B<0$ or $\tilde{\chi}_A\tilde{\chi}_B<-1$. Under this condition, the instability is oscillatory if $|\tilde{\chi}_A-\tilde{\chi}_B|<2$ and stationary otherwise.

\subsubsection{Short-wave oscillatory instabilities}

\label{sec:SOI_general}

The previous expansion of~\eqref{seq:M_minus_lambda} for vanishing $q$ cannot capture the onset of short-wave instabilities for which $\Re(\lambda)(q)$ is positive only beyond $q_{\rm min} > 0$.
Instead, we show here how this analysis can be achieved without explicitly solving the quartic equation $|\underline{\mathbf{M}}-\tilde{\lambda}\underline{\mathbf{I}}|=0$ by use of the Routh-Hurwitz stability criterion.
Below, we first give a brief introduction to the Routh-Hurwitz stability criterion (see Chapter \uppercase\expandafter{\romannumeral5} of Ref.~\cite{gantmacher2005applications} for a detailed presentation), and then  use it to obtain the conditions for the emergence of SOI. 

We consider a $N$ dimensional dynamical system linearized around a given stationary state. The corresponding growth rate is then given as a root of a degree $N$ polynomial $P(\lambda) = a_0\lambda^N+a_1 \lambda^{N-1} +\dots + a_{N-1} \lambda + a_N$.
The steady state will thus be stable if and only if all roots of $P(\lambda)=0$ lie in the left half of the complex plane (i.e.\ if they all have a negative real part).
The Routh-Hurwitz criterion~\cite{gantmacher2005applications} states that a necessary and sufficient condition for $\Re(\lambda) < 0$ is $\Delta_j>0$ $\forall j=1,\dots,N$, where the $\Delta_j$ are the so-called Hurwitz determinants~\cite{gantmacher2005applications}.
For the case $N = 4$ of interest here, we have
\begin{equation*} \label{seq:Hurwitz-det}
    \Delta_1 = a_1, \quad
    \Delta_2
    = a_1 a_2 - a_0 a_3, \quad
    \Delta_3
    = a_3\Delta_2 - a_1^2 a_4,\quad
    \Delta_4
    = a_4 \Delta_3,
\end{equation*}
where $a_0=1$
and the other coefficients read from~\eqref{seq:M_minus_lambda}:
\begin{equation*}\label{seq:a_1_4}
        a_1 = \gamma_A + \gamma_B,\quad
        a_2 = \gamma_A\gamma_B + \frac{1}{2}\left(\chi_{B}+\sigma_v^2\chi_{A}\right), \quad
        a_3 = \frac{1}{2}\left(\gamma_A\chi_{B}+\sigma_v^2\gamma_B\chi_{A}\right) , \quad
        a_4 = \frac{\sigma_v^2}{4} (\chi_{A}\chi_{B}- \chi).
\end{equation*}
Due to the dependence of $\gamma$ on $q$, $a_{1, 2, 3}$ are $q$-dependent while $a_4$ has no dependency on $q$.

A corollary of the Routh-Hurwitz criterion is that $a_j>0$ for all $j=1,\dots,N$ is a necessary condition for stability, which conversely means that $a_j < 0$ for at least one value of $j$ is a sufficient condition for instability.
As $\gamma_S \sim q^{-1} > 0$ for vanishing $q$, it immediately follows that $a_{1,2}(q) > 0$ for $q \to 0$ such that the sufficient conditions for the emergence of long-wave instability in the linear system \eqref{seq:u_q_t} are $a_3(0^+)<0$ or $a_4<0$, which are equivalent to that derived
in Sec.~\ref{sec:LI}.
Conversely, if
\begin{equation} \label{seq:condi-no-long-wave-inst}
    \chi_A + \sigma \chi_B > 0 \quad \mathrm{and} \quad \chi_A\chi_B > \chi,
\end{equation}
we have $\Delta_j(0^+)>0$, $a_j(0^+)>0$ for $j=1, 2, 3$ and $a_4>0$, so that no long-wave instability can arise.

Under the condition~\eqref{seq:condi-no-long-wave-inst} and since $\gamma_S > 0$, the short-wave instability can occur if and only if $\Delta_2$ or $\Delta_3$ change sign at some $q^*\in(0, 1)$.
Moreover, $\Delta_2 = 0$ implies that $\Delta_3 = - a_1^2 a_4$ while $\Delta_{2, 3}$ and $a_{4}$ are all positive for $q \to 0$, it is clear that if $\Delta_2$ becomes negative at $q = q^*$, then $\Delta_3$ must have already changed sign at $q< q^*$.
Hence, in what follows we only need to study the signs of $\Delta_3$ as function of $q$.

As the quartic equation~\eqref{seq:M_minus_lambda} includes real coefficients $a_j$, 
it can possess either four real roots, two real and a pair of complex conjugate roots, or two pairs of complex conjugate roots. The number of roots with positive real part is furthermore given by the number of sign changes between consecutive elements of the Routh array \cite{gantmacher2005applications}: $[1, \Delta_1, {\Delta_2}/{\Delta_1}, \Delta_3/\Delta_2, a_4]$. 
Therefore, a change in the sign of $\Delta_3$ only implies two sign changes in the Routh array, and thus the presence of two roots with positive real part. 
Besides, setting $\Delta_3 = 0$ gives the relation $a_3 \Delta_2 = a_1^2 a_4 > 0$, while the coefficient $a_3$ can be expressed in terms of the eigenvalues of $\underline{\bf M}$ as $a_3 = -\sum_{i=1}^4 \prod_{j\ne i} \tilde{\lambda}_j$.
Hence, if two real eigenvalues simultaneously change sign at $q = q^*$, then $a_3(q^*) = 0$ such that $\Delta_3(q^*) \ne 0$.
We thus conclude that $\Delta_3$ can only change sign when two complex conjugate eigenvalues have their real part become positive, implying that the steady state undergoes SOI.

Therefore, SOI will arise if $a_{3}(0^+)>0$, $a_4>0$, and $\Delta_3$ becomes negative for $q \in (0,1)$. 
We first write
$\Delta_3 = \Delta_3^{[0]} +\Delta_3^{[1]} +\Delta_3^{[2]}$, where
\begin{equation*}
    \Delta_3^{[0]}
    \equiv \frac{\gamma_A\gamma_B}{4} (\sigma_v^2\chi_{A}-\chi_{B})^2, \quad
    \Delta_3^{[1]} \equiv \frac{\sigma_v^2}{4}(\gamma_A+\gamma_B)^2\chi, \quad
    \Delta_3^{[2]}
    \equiv \gamma_A\gamma_B a_1 a_3.
\end{equation*}
Since $\Delta_3^{[0]}>0$, $\Delta_3(q)$ can only change sign for $q \in(0, 1)$ if either $\Delta_3^{[1]}$ or $\Delta_3^{[2]}$ become negative. 
It is straightforward that $\Delta_3^{[1]}<0$ is equivalent to $\chi<0$. 
On the other hand, $\Delta_3^{[2]}<0$ is equivalent to $a_3 < 0$,
while the precondition for $a_3$ to change sign when $q$ is increased is given by $\chi_A\chi_B<0$.
Meanwhile, to guarantee $a_4\propto \chi_A\chi_B-\chi>0$, it is necessary to have $\chi<\chi_A\chi_B< 0$. 
Therefore, in both cases $\chi<0$ is required, and is thus a prerequisite for SOI.

We can further show that for general $\sigma_v$, SOI is forbidden at vanishing $\mathrm{Pe}$, in which case we find
\begin{equation*}
    \Delta_3^{[2]}\underset{\rm Pe \to 0}{\simeq} \frac{1}{\mathrm{Pe}^4q^4}\left(\chi_A+\frac{\chi_B}{\sigma_v^2}\right)>0, \qquad q\in(0, 1),
\end{equation*}
given that $a_3(0^+)\propto\chi_A+\chi_B/\sigma_v^2>0$. 
Moreover, as for $\mathrm{Pe}=0^+$ we have $\Delta_3^{[2]}\sim \mathrm{Pe}^{-4}\gg |\Delta_3^{[1]}|\sim \mathrm{Pe}^{-2}$, we conclude that $\Delta_3$ must remain positive such that SOI cannot arise in the limit $\mathrm{Pe} \to 0$.

For $\mathrm{Pe}\sim \mathcal{O}(1)$ and in the limit of small $q$, we find $\gamma_A(q)\sim
\gamma_B(q)$. With this approximation, we find that $\Delta_3(q)$ will become negative at some $q\in(0, 1)$ once
\begin{equation} \label{eq:condi_SOI_small_q}
    \gamma_B^2 (\sigma_v^2\chi_{A} + \chi_{B}) + \sigma_v^2\chi + \frac{1}{4}(\sigma_v^2\chi_A-\chi_B)^2 < 0.
\end{equation}
This condition reduces to Eq.~\eqref{eq:condi_pos_lambda_neg_alpha} when $\sigma_v=1$.
Since $\sigma_v^2\chi_A+\chi_B\propto a_3(0^+)>0$ and $\gamma_B\sim (\mathrm{Pe}q)^{-1}$ for small $q$, to allow the condition \eqref{eq:condi_SOI_small_q} to be true for $q\in (0, 1)$, we need large enough $\mathrm{Pe}$ and $|\chi|$ provided $\chi < 0$.

\subsection{Linear stability diagram} \label{sec:linear-stability-diagram}

Here, we provide details on how the linear stability diagrams of Figs.~1 and 2 of the main text were obtained.
In all cases, the conditions for LSI and LOI are known analytically, as shown in
Sec.~\ref{sec:LI}.
At larger $\rm Pe$ we instead have to consider the possibility for SOI (Fig.~1(e, f) of the main text). 
At fixed parameters satisfying~\eqref{seq:condi-no-long-wave-inst} such that long-wave instabilities are forbidden, SOI arise if the parameter $\Delta_3$ defined in Sec.~\ref{sec:SOI_general} becomes negative with increasing $q$.

For the linear stability diagram in composition plane (inset of Fig.~2(a)), although the nominal couplings $\{\eta^0_{SS'}\}$ of the microscopic model are fixed, the mean-field couplings $\{\bar{\eta}_{SS'}\}$ entering into the linear stability analysis depend on the composition $(\bar{\rho}_A,\bar{\rho}_B)$. 
Scanning the $(\bar{\rho}_A,\bar{\rho}_B)$ plane with a fine resolution, we thus determine the corresponding linear stability diagram from the aforementioned criteria that decide the nature of the instabilities, leading to the linear stability diagram such as the inset of Fig.~2(a) of the main text.

We moreover checked that a direct numerical computation of the eigenvalues of the matrix $\underline{\mathbf{M}}(q)$ leads to the same results as those given by the Routh-Hurwitz criterion.

\section{Coarsening dynamics}

\begin{figure}
	\includegraphics[width=0.5\columnwidth]{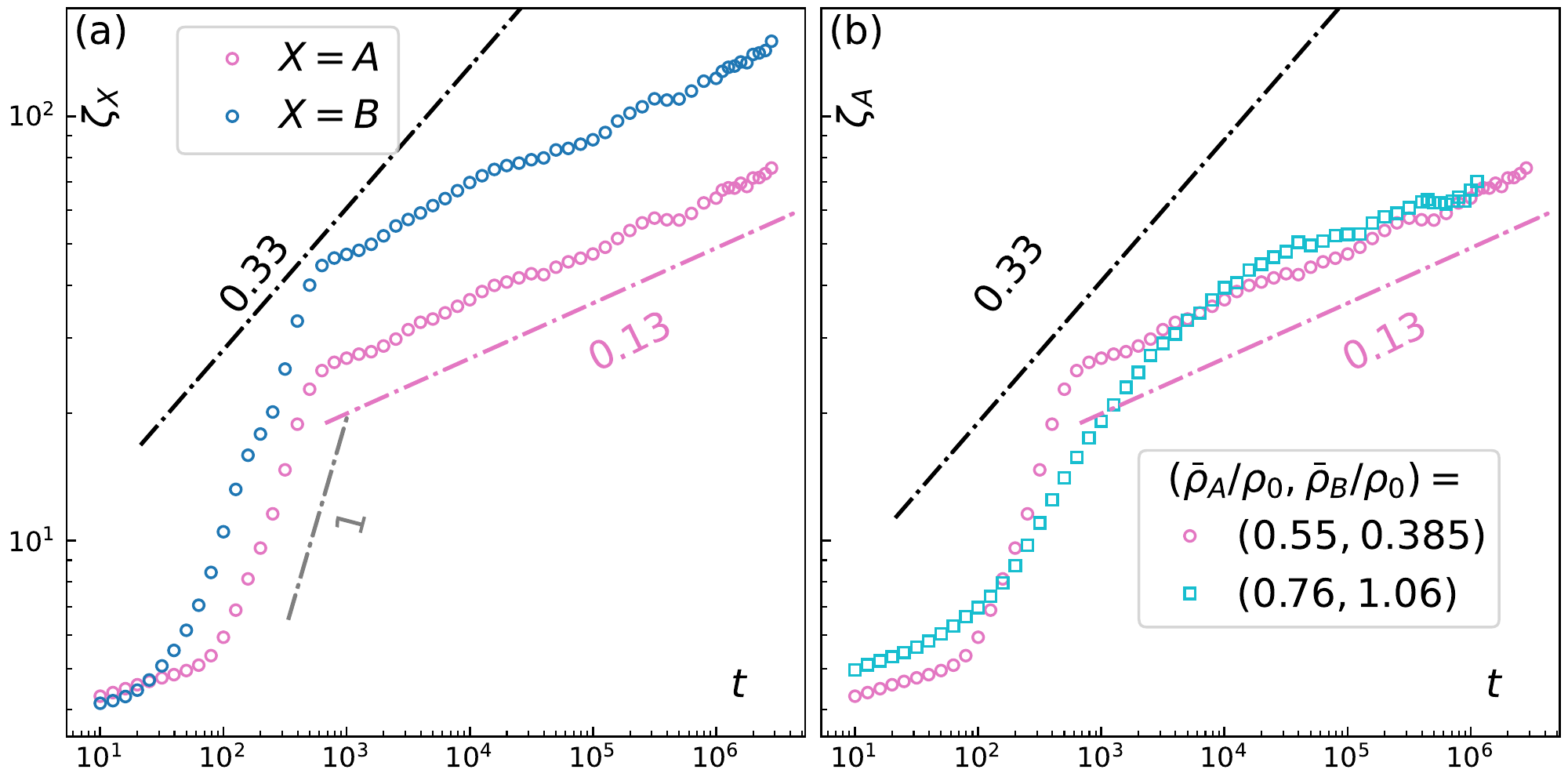}
	\caption{(a) The growth of domain sizes $\zeta_A$ and $\zeta_B$ in the CCB-gas coexistence phase with $(\bar\rho_A/\rho_0=0.55, \bar\rho_B/\rho_0=0.385)$. (b) $\zeta_A$ vs. $t$ for two different compositions in the CCB-gas coexistence phase. The pink curve here corresponds to the pink curve in Fig.~3(e) of the main text, while the composition for the cyan curve here is marked by the cyan square in Fig.~2(a) of the main text.
	Other parameters: $L_x=L_y=1280$, $\eta_{AA}^0=\eta_{BB}^0=-2$, $\eta_{AB}^0=-\eta_{BA}^0=0.5$, $D_r=0.1$, $\rho_0=10$.
	 }
	\label{FS1}
\end{figure}

To quantify coarsening, we defined the typical domain sizes $\zeta_{A,B}(t)$ as
\begin{equation} \label{eq:R_X_t}
    \zeta_X (t)=2\pi \int \mathrm{d} q\, \tilde{S}_X(q,t)\bigg/\int \mathrm{d}q\, q \tilde{S}_X(q,t),
\end{equation}
where $X\in \{A,B\}$ and 
$S_X(\mathbf{q},t)=\langle\phi_X(\mathbf{q},t)\phi_X(-\mathbf{q},t) \rangle$
is the structure factor for the density fields of species $X$,
while $\tilde{S}_X(q,t)\equiv \langle S_X(\mathbf{q}, t)\rangle_{|\mathbf{q}|=q}$ denotes the corresponding average over all orientations of $\bf q$.
As shown in Fig.~\ref{FS1}(a), $\zeta_A(t)$ and $\zeta_B(t)$ have similar growth behaviors, such that in the main text we only show the domain size of the majority species.
We moreover have studied the coarsening of CCB domains at higher density marked by the cyan solid square in Fig.~2(a) of the main text. At higher density (cyan curve in Fig.~\ref{FS1}(b)), the scaling regime is reached much later than the lower density case (pink curve in Fig.~\ref{FS1}(b)).
This is because the scaling regime should be reached after the mean distance among liquid domains becomes large enough. For higher densities, there would be more droplets at early stages, thus we need to wait longer until the mean distance among liquids domains becomes large enough. Nevertheless, at late stages, the coarsening behavior at higher densities is similar to that in the lower density regime. In the latter case, the particle-based simulation is much less computationally intensive, allowing us to reach larger time scales.

\section{Parameters for the figures in the main text and supplemental movies.}

Here, we list the parameters used for the figures in the main text.

\begin{itemize}
\item Fig.~1: In panel (a--d), the simulations are performed in square boxes with $L_x=40$ and initialized from randomly with $\bar{\rho}_A=\bar{\rho}_B=\rho_0=80$ and $D_r=0.1$. Other parameters: (a) $\eta^0_{AA}=\eta^0_{BB}=-2$, $\eta_{AB}^0=-\eta_{BA}^0=0.1$; (b) $\eta^0_{AA}=0$, $\eta^0_{BB}=-2$, $\eta_{AB}^0=-\eta_{BA}^0=0.1$; (c) $\eta^0_{AA}=\eta^0_{BB}=0$, $\eta_{AB}^0=-\eta_{BA}^0=0.8$; (d) $\eta^0_{AA}=\eta^0_{BB}=0$, $\eta_{AB}^0=-\eta_{BA}^0=1.8$.

\item Fig.~2: The simulations are performed in square boxes with $L_x=40$ in (a) and $L_x=160$ in (b--d). The compositions $(\bar{\rho}_A/\rho_0,\bar{\rho}_B/\rho_0)$ for (b--d) are $(2, 2)$, $(0.75, 1.25)$ and $(1, 0.75)$ respectively. Other parameters are: $\eta_{AA}^0 = \eta_{BB}^0 = -2$, $\eta_{AB}^0 = -\eta_{BA}^0 = 0.5$, $D_r=0.1$, $\rho_0=20$.

\item Fig.~3: In (a--c) simulation were performed in a rectangular box with $L_x=3L_y=960$, at three compositions $(\bar{\rho}_A/\rho_0,\bar{\rho}_B/\rho_0)=(0.7625, 1.05675)$, $(0.8625, 1.336)$ and $(0.9597, 1.6152)$. In (e), the simulations were run in a square box of size $L_x=1280$, at three compositions $(\bar{\rho}_A/\rho_0,\bar{\rho}_B/\rho_0)$: SPS $(0.3, 0.6)$, CCB-gas $(0.55, 0.385)$ and CCB $(2, 2)$. Other parameters: $\eta_{AA}^0 = \eta_{BB}^0 = -2$, $\eta_{AB}^0 = -\eta_{BA}^0 = 0.5$, $D_r=0.1$, $\rho_0=10$.
\end{itemize}

Below, we describe the videos available as supplemental material.

\begin{itemize}
\item \path{SMov1}. Corresponding to Fig.~1(a--d) of the main text, where four types of patterns in square boxes with $L_x=40$ are shown after reaching the steady state.
\item \path{SMov2}. Corresponding to Figs.~2(b--d) of the main text, where we show that at large system size the regular travelling bands and spirals are superseded by CCB-gas coexisting phase, while the pure CCB phase remains intact.
\item \path{SMov3}. Corresponding to Fig.~3(d) of the main text, the close-up of CCB-gas interface in the steady state.
\end{itemize}

\bibliography{NRQS_supp.bib}

\date{\today}